\documentclass[sigconf]{acmart}

\copyrightyear{2026}
\acmYear{2026}
\setcopyright{cc}
\setcctype{by}
\acmConference[ICAIF '26]{7th ACM International Conference on AI in Finance}{November 14--17, 2026}{Milan, AA, Italy}
\acmBooktitle{7th ACM International Conference on AI in Finance (ICAIF '26), November 14--17, 2026, Milan, AA, Italy}
\acmDOI{10.1145/3858063.3858902}
\acmISBN{979-8-4007-3087-0/2026/11}
\usepackage{amsmath}
\usepackage{booktabs}
\usepackage{graphicx}
\usepackage{placeins}

\newcommand{\PaperNCalls}{552}
\newcommand{\PaperNTickers}{24}
\newcommand{\PaperRetention}{26.4\%}

\newcommand{\PaperEventPanelN}{2,094}
\newcommand{\PaperFirstYearN}{2}
\newcommand{\PaperLaterYearMinN}{87}
\newcommand{\PaperLaterYearMaxN}{95}
\newcommand{\WorkedTicker}{MSFT}
\newcommand{\WorkedCallDate}{27 July 2021}
\newcommand{\WorkedCallStart}{17:00}
\newcommand{\WorkedAudioProxy}{18:00:20}
\newcommand{\WorkedReadyProxy}{18:15:20}
\newcommand{\WorkedAnchor}{09:35}
\newcommand{\WorkedLegacyReturn}{-0.984}
\newcommand{\WorkedPreAnchorReturn}{-0.562}
\newcommand{\WorkedPostProxyReturn}{-0.422}
\newcommand{\DECCommonN}{1,710}
\newcommand{\DECBeforeMarketN}{1,191}
\newcommand{\DECNumericTargetChange}{69.6\%}
\newcommand{\DECThreeFixedQuintileMove}{43.5\%}

\newcommand{\DECFifteenFixedQuintileMove}{23.7\%}
\newcommand{\DECThreeMedianRankDisplacement}{12.5}

\newcommand{\DECFifteenMedianRankDisplacement}{6.1}

\newcommand{\QCMSTargetQuintileMove}{68.5\%}

\newcommand{\LegacySourceSpearman}{0.506}
\newcommand{\LegacySourceMeanAbs}{3.427}

\newcommand{\LegacySourceSignFlip}{32.1\%}
\newcommand{\LegacySourceRelativeMove}{64.3\%}

\newcommand{\SourceCompletionSpearman}{0.900}
\newcommand{\SourceCompletionMeanAbs}{0.623}
\newcommand{\SourceCompletionMedianAbs}{0.179}
\newcommand{\SourceCompletionSignFlip}{9.4\%}
\newcommand{\SourceCompletionChangedAnchorN}{311}
\newcommand{\SourceCompletionChangedAnchorRate}{56.3\%}
\newcommand{\SourceCompletionEqualAnchorN}{241}
\newcommand{\SourceCompletionAtRiskSignFlip}{16.7\%}
\newcommand{\SourceCompletionAtRiskRelativeMove}{48.2\%}
\newcommand{\SourceCompletionAtRiskFixedMove}{50.2\%}
\newcommand{\SourceCompletionEqualAnchorRelativeMove}{6.6\%}
\newcommand{\SourceCompletionRelativeMove}{30.1\%}
\newcommand{\SourceCompletionFixedMove}{28.3\%}
\newcommand{\SourceCompletionSignFirmLow}{5.6\%}
\newcommand{\SourceCompletionSignFirmHigh}{13.6\%}
\newcommand{\SourceCompletionSignWeekLow}{6.9\%}
\newcommand{\SourceCompletionSignWeekHigh}{12.1\%}
\newcommand{\SourceCompletionRelativeFirmLow}{20.6\%}
\newcommand{\SourceCompletionRelativeFirmHigh}{40.2\%}
\newcommand{\SourceCompletionRelativeWeekLow}{25.3\%}
\newcommand{\SourceCompletionRelativeWeekHigh}{35.5\%}
\newcommand{\SourceCompletionFixedFirmLow}{17.7\%}
\newcommand{\SourceCompletionFixedFirmHigh}{38.1\%}
\newcommand{\SourceCompletionFixedWeekLow}{23.4\%}
\newcommand{\SourceCompletionFixedWeekHigh}{31.9\%}
\newcommand{\AfterCloseSourceCompletionN}{206}
\newcommand{\AfterCloseSourceCompletionRelativeMove}{6.3\%}
\newcommand{\DECThreeSameBestIssuer}{67.1\%}
\newcommand{\DECThreeSameBestQuarter}{72.1\%}
\newcommand{\DECSevenSameBestIssuer}{95.0\%}
\newcommand{\DECSevenSameBestQuarter}{83.3\%}
\newcommand{\DECFifteenSameBestIssuer}{86.1\%}
\newcommand{\DECFifteenSameBestQuarter}{80.4\%}
\newcommand{\DECThreeAnyInversionIssuer}{43.1\%}
\newcommand{\DECThreeAnyInversionQuarter}{45.0\%}

\newcommand{\ValenceLegacyBeta}{+1.122}
\newcommand{\ValenceSourceBeta}{-0.135}
\newcommand{\ValenceAvailableBeta}{-0.233}
\newcommand{\ValenceLegacySourceDelta}{+1.256}
\newcommand{\ValenceSourceAvailableDelta}{+0.098}
\newcommand{\ValencePairedBeta}{+1.355}
\newcommand{\ValenceFirmCILow}{+0.619}
\newcommand{\ValenceFirmCIHigh}{+2.098}
\newcommand{\ValenceWeekCILow}{+0.645}
\newcommand{\ValenceWeekCIHigh}{+2.027}

\newcommand{\ValenceHolmP}{0.0024}
\newcommand{\ValenceExactBeta}{-0.226}

\newcommand{\AllowanceSignFlipMin}{6.5\%}
\newcommand{\AllowanceSignFlipMax}{15.0\%}
\newcommand{\AllowanceRelativeMoveMin}{23.9\%}
\newcommand{\AllowanceRelativeMoveMax}{39.9\%}
\newcommand{\AllowanceFixedAFifteenMoveMin}{21.0\%}
\newcommand{\AllowanceFixedAFifteenMoveMax}{35.9\%}

\title[When Does ``Post-Call'' Begin?]{When Does ``Post-Call'' Begin? An Availability Contract for Earnings-Call NLP Benchmarks}

\author{Xiaoyu Ma}
\affiliation{%
  \institution{National University of Singapore}
  \department{Asian Institute of Digital Finance}
  \city{}
  \country{Singapore}
}
\email{xiaoyu\_ma@u.nus.edu}

\author{Kelvin J.L. Koa}
\authornote{Corresponding author.}
\affiliation{%
  \institution{National University of Singapore}
  \department{Asian Institute of Digital Finance}
  \city{}
  \country{Singapore}
}
\email{kelvin.koa@u.nus.edu}

\author{Yuncong Liu}
\affiliation{%
  \institution{National University of Singapore}
  \department{Asian Institute of Digital Finance}
  \city{}
  \country{Singapore}
}
\email{yuncongliu@u.nus.edu}

\author{Ke-Wei Huang}
\affiliation{%
  \institution{National University of Singapore}
  \department{Asian Institute of Digital Finance}
  \city{}
  \country{Singapore}
}
\email{dishkw@nus.edu.sg}

\renewcommand{\shortauthors}{Ma et al.}
\begin{abstract}

Corporate financial disclosures often unfold over time, with announcements, prepared remarks, and discussion spanning multiple high-frequency return intervals. Financial NLP benchmarks can therefore suffer from a within-event timing mismatch when the prediction target begins before the textual or audio input used by a model could be complete. Chronological train--test splits do not prevent this mismatch, because it occurs inside individual events rather than across samples. We introduce an \emph{Availability Contract}, which requires a benchmark to state the input boundary, evidence for input availability, the outcome start, and the claims supported by that evidence. We apply this framework to \PaperNCalls{} earnings calls from \PaperNTickers{} firms. For each call, we hold the return endpoint fixed and compare the previous close, a documentary call-start proxy, and a completion proxy based on reconstructed audio end plus a stipulated processing allowance. At this baseline, moving from the call-start proxy to the completion proxy flips return signs for \SourceCompletionSignFlip{} of calls. Holding completion-clock cutpoints fixed, \SourceCompletionFixedMove{} change quintile. In an independent release-session analysis of a public earnings-call volatility benchmark, shifting before-market target windows forward by one trading session raises three-day mean squared error by 18.1--21.4\% across four frozen prediction streams, while their point rankings remain unchanged. Together, the results show that labels and losses can depend materially on the declared outcome clock even when model rankings appear stable. The \emph{Availability Contract} makes this assumption explicit and ties post-input claims to the available timing evidence.

\end{abstract}



\begin{document}
\maketitle
\raggedbottom

\section{Introduction}

Financial corporate events unfold over time. An earnings call may include an
hour of prepared remarks and Q\&A while stock prices continue to change.
Press conferences and other sequential disclosures create the same benchmark
problem. Let $X_i$ denote the model's textual or audio input for event $i$ and
$Y_i$ its target stock return. Post-input prediction requires the return
window to begin at a price observed after the complete $X_i$ could have been
constructed. An earlier window includes price movement during disclosure:
it may measure a meaningful event-response association, but not prediction
after the declared input becomes available.


A chronological train--test split excludes future training events, but not
this \emph{within-event} mismatch. Unlike conventional train--test leakage,
the issue concerns the \emph{representation boundary}---the information
included in the input---and the \emph{label start}---the price from which
the outcome is measured.

These choices affect reproducibility in financial AI. Identical models, data,
and splits can yield different labels when disclosure times, input completion,
or timestamp-to-price mappings differ. Such assumptions may be buried in
preprocessing code. Interpreting an event-window benchmark as post-input
evaluation therefore requires reproducible event timing as well as model code.

\begin{table*}[t]
  \caption{Claim ladder for availability evidence.}
  \label{tab:claim-ladder}
  \centering
  \small
  \begin{tabular}{@{}p{1.35in}p{5.45in}@{}}
    \toprule
    Claim & Permitted interpretation \\
    \midrule
    Event-window association & The return may span the event; interpretation is
      an event-window association, not post-input attribution. \\
    Proxy-aligned association & A stated completion proxy precedes label start;
      interpretation is conditional on the proxy. \\
    Verified post-input & Observed input completion, or a defensible upper
      bound, precedes label start. \\
    Deployability & Point-in-time input, model, order, and execution clocks are
      observed or conservatively bounded, and costs are modeled. \\
    \bottomrule
  \end{tabular}
\end{table*}

We introduce an \emph{Availability Contract} to make these assumptions
explicit. The contract records the information included in the model input,
the evidence used to determine when that input is complete, any processing
allowance, the rule mapping completion to a market-price anchor, the outcome
endpoint, and the interpretation supported by the available timing evidence.
It does not impose one universally correct event window. A return beginning
before event completion may be appropriate for studying the total market
response, whereas a return beginning after completion is more appropriate for
evaluating prediction using the complete representation. The contract requires
researchers to distinguish these estimands rather than describe both as
post-input prediction.

We study this issue using earnings calls, whose sequential structure makes the
timing distinction especially visible. Prior research shows
that earnings-call language and audio are associated with market outcomes
\cite{matsumoto2011conference,sawhney2020voltage}. However, some benchmarks
combine complete-call representations with returns beginning at or before the
start of the call
\cite{sawhney2020voltage,voltageartifact,amazon2017call,sang2022dialoguegat}.
Such designs may be appropriate for event-response analysis, but additional
clock evidence is required before interpreting them as evidence of forecast accuracy.

For each call, we construct returns with the same endpoint but three starting
prices. The legacy clock $L$ begins at the preceding-session close and captures
a broad event-spanning return. Clock $S$ uses the close of the first
regular-session five-minute bar starting strictly after a documentary
source-start timestamp. Clock $A$ applies the same rule after the reconstructed
audio end plus a stipulated 15-minute allowance. Each selected bar closes five
trading minutes after its start, so neither anchor is an instantaneous price at
the proxy timestamp. Moving from $L$ to $S$ removes price movement before the
documentary source-start anchor. Moving from $S$ to $A$ additionally removes
the reconstructed audio-duration interval, the allowance, and the subsequent
bar-mapping interval. Because the sample and endpoint remain fixed, target
differences arise from the intervals removed by changing the starting price.
Across \PaperNCalls{} earnings calls from \PaperNTickers{} firms, at the
stipulated 15-minute baseline, moving from $S$ to $A$ flips the return sign for
\SourceCompletionSignFlip{} of calls. Holding $A$-clock cutpoints fixed,
\SourceCompletionFixedMove{} change quintile; recomputing categories separately
under each clock yields \SourceCompletionRelativeMove{}.

We complement this audit with a separate sensitivity analysis using the public
Dense Earnings Call (DEC) benchmark introduced by \citet{yu2025identity}.
Moving the targets of before-market events forward by one
trading session increases three-day mean squared error by 18.1--21.4\% across
four fixed prediction streams, although their rankings remain unchanged. This
shows that an unchanged leaderboard does not imply that target values or
reported predictive losses are robust to timing choices.

Our contributions are threefold. First, we introduce the Availability Contract
as a representation-specific reporting framework for reproducible financial
NLP benchmarks. Second, we provide a paired three-clock diagnostic that
separates event-spanning, call-start-aligned, and completion-aligned return
windows while holding the sample and endpoint fixed. Third, we show that timing
choices can reverse return signs, reassign categorical labels, and materially
change reported losses even when model rankings remain stable.

\begin{figure*}[!t]
  \centering
  \includegraphics[width=\textwidth]{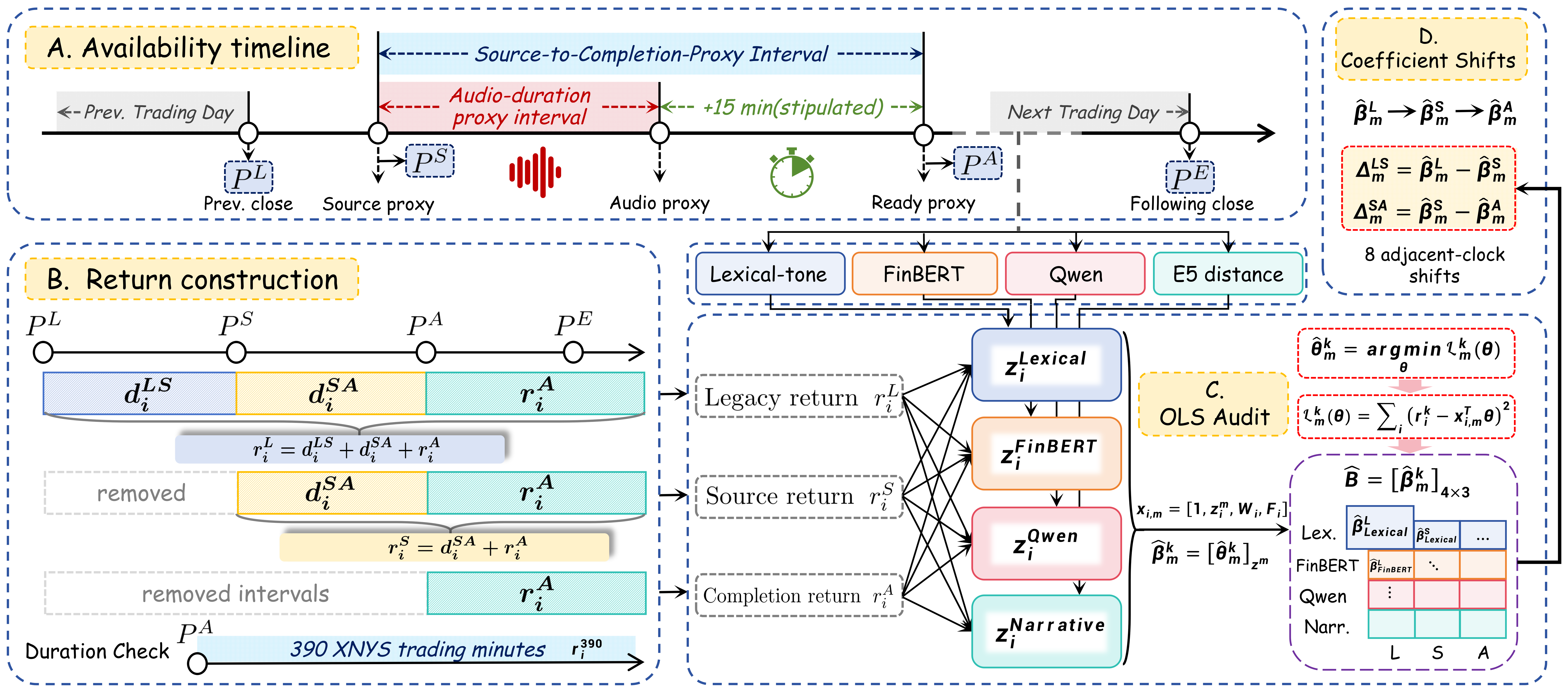}
  \caption{Availability contract and paired three-clock audit.
  Panel A maps source and stipulated completion proxies to non-executable
  price anchors; transcript and model readiness remain unobserved.
  Panel B constructs three common-endpoint returns and their removed
  intervals. Panel C crosses four retrospective representations with three
  targets in 12 common-sample OLS fits. Panel D reports eight adjacent-clock
  coefficient contrasts. These contrasts are associational, $r_i^{390}$ is a
  separate duration check, and the timeline is not to scale.}
  \Description{Four panels show the proxy timeline, return decomposition,
  common-sample OLS audit, and adjacent-clock coefficient contrasts.
  The representations are computed retrospectively, and the price anchors
  are labels rather than execution prices.}
  \label{fig:contract}
\end{figure*}
\section{Related Work and Evaluation Context}

\subsection{Earnings calls as information events}

Financial language contains information beyond accounting numbers
\cite{tetlock2008words}. Earnings-call Q\&A adds to the presentation
\cite{matsumoto2011conference}; non-answers and linguistic patterns reveal
managerial behavior
\cite{gow2021nonanswers,larcker2012deceptive,chen2025conversation}; and
question--answer similarity has been related to returns
\cite{hynes2026evasion}. These findings motivate complete-interaction
representations, which cannot exist when the call begins. Economic relevance
and temporal validity are separate questions.

\subsection{Financial-NLP representations and benchmarks}

Financial modeling uses finance-specific dictionaries and FinBERT
\cite{loughran2011liability,araci2019finbert}, retrieval-augmented price--news
representations for portfolio optimization \cite{koa2026radar}, and
multi-agent symbolic discovery for interpretable valuation
\cite{koa2026mufasa}. Controlled synthetic benchmarks
diagnose forecasting performance under explicit financial mechanisms
\cite{sun2026finstressts}. Earnings-call research spans long-document
prediction \cite{koval2023earnings}, aligned multimodal data
\cite{li2020maec}, text--audio fusion \cite{sawhney2020voltage}, broader
multimodal benchmarks \cite{shu2026fincall}, and annotated Q\&A quality
\cite{pardawala2024subjective}.

Recent ICAIF work on earnings-release structure separates announcement-day
explanatory power from a post-open trading test \cite{wu2025press}. The
Availability Contract extends this reporting principle from release timing to
representation-specific completion.
Transcript models may also load on company identity and prior volatility rather
than call-specific semantics \cite{yu2025identity}. Yu et al.\ additionally show
that day indexing in the EC and MAEC earnings-call benchmarks can select the
wrong first post-earnings day when release timing is omitted. Event-date correction and chronological splitting therefore remain
necessary. Related text-to-decision work separates directional intent from
execution choices \cite{liu2026missing}; our contract additionally binds
representation completion to a price anchor and endpoint.

\subsection{Event windows, temporal validity, and multiplicity}

Classic event studies measure reactions around dated events
\cite{brown1985daily,mackinlay1997event,kolari2010event}. Announcement-clock
research shows that informative windows depend on release time
\cite{berkman2009eventday,das2021eventwindows}; at segment resolution, Oh links
timestamped call speech to subsequent quote returns \cite{oh2026price}. We ask what evidence permits a benchmark label to begin after its declared
representation is available. Predictive evaluation requires excluding unavailable inputs
\cite{kaufman2012leakage}.

Random splits can exaggerate generalization under distribution shift
\cite{sogaard2021random}, and pretrained models can encode post-cutoff facts
\cite{liu2026exante,gao2025lookahead}. Point-in-time checkpoints address model
knowledge \cite{kelly2026pointintime}, not a label that begins before its call
input is complete. We therefore treat label start as benchmark schema.
Table~\ref{tab:claim-ladder} lists these claims.

Repeated predictor search also makes isolated significance difficult to
interpret \cite{harvey2016crosssection}. Multiplicity adjustment addresses
the resulting statistical problem, whereas the contract addresses the
definition of the outcome being tested.

\section{Availability-Aware Paired Audit}

\subsection{Availability Contract and instantiated proxy clocks}

An Availability Contract records nine fields: permitted claim $q$; input boundary $B$;
source-to-completion-proxy rule $G$; allowance $\ell$; checkpoint record $K$,
including its identity, knowledge boundary, and point-in-time availability;
inference-completion evidence $M$; price-anchor rule $\mathcal{A}$; endpoint
$E$; and sample rule $\mathcal{S}$. Label start applies $\mathcal{A}$ to
$G(B)+\ell$ under the stated market calendar. The contract specifies a
reproducible window whose interpretation follows the recorded evidence for
$G$, $K$, and $M$: a proxy-aligned
association, a verified post-input evaluation, or a deployment claim.

Let $t_i^{\mathrm{input}}$, $t_i^{\mathrm{checkpoint}}$,
$t_i^{\mathrm{model}}$, $t_i^{\mathrm{order}}$, and
$t_i^{\mathrm{exec}}$ denote the complete-input, checkpoint-available,
model-ready, order, and execution times for call $i$. A deployment claim
requires label start to follow every clock relevant to the claimed workflow,
with a checkpoint whose knowledge boundary precedes the forecast. These clocks
are unobserved here. Figure~\ref{fig:contract} therefore shows only the
documentary source timestamp, reconstructed audio-duration proxy, stipulated
allowance, and two derived price anchors. The four representations are
constructed retrospectively and connected to these clocks through the
offline audit.

For each call, we reconstruct an audio-duration proxy as
\begin{equation}
 t_i^{\mathrm{audio}}
 =
 t_i^{\mathrm{source\ start}}
 +
 \max\!\left(
 d_i^{\mathrm{MP3}},
 e_i^{\mathrm{alignment}}
 \right),
 \label{eq:audio}
\end{equation}
where $d_i^{\mathrm{MP3}}$ is the decoded MP3 duration and
$e_i^{\mathrm{alignment}}$ is the elapsed time of the final finite
sentence-alignment endpoint, both measured relative to the documentary source
timestamp. Taking the maximum makes the proxy no earlier than either recorded
duration estimate. These measurements reconstruct an audio-duration boundary
relative to the documentary source timestamp.

We then stipulate
\begin{equation}
 t_i^{\mathrm{ready\ proxy}}
 =
 t_i^{\mathrm{audio}}+\ell,
 \qquad
 \ell=15\text{ minutes}.
 \label{eq:ready_proxy}
\end{equation}
This ready proxy supplies the study's completion-proxy clock. Let
$\mathcal{U}_{\mathrm{XNYS}}$ denote regular-session five-minute bar-start
timestamps on the New York Stock Exchange (XNYS) calendar. We define two intermediate
label-price anchors:
\begin{equation}
 \begin{aligned}
 u_i^S
 &=
 \min\left\{
 u\in\mathcal{U}_{\mathrm{XNYS}}:
 u>t_i^{\mathrm{source\ start}}
 \right\},
 &
 P_i^S
 &=
 \mathrm{Close}_i(u_i^S),
 \\
 u_i^A
 &=
 \min\left\{
 u\in\mathcal{U}_{\mathrm{XNYS}}:
 u>t_i^{\mathrm{ready\ proxy}}
 \right\},
 &
 P_i^A
 &=
 \mathrm{Close}_i(u_i^A).
 \end{aligned}
 \label{eq:anchors}
\end{equation}

Each selected bar starts strictly after its corresponding proxy and closes
five trading minutes later. $P_i^S$ follows the documentary source-start proxy,
while $P_i^A$ incorporates the reconstructed duration and stipulated allowance.
Both are bar-close prices used to construct the return labels. XNYS sessions,
including early closes, are implemented with
\texttt{exchange\_calendars} 4.5.6. These rules define a reproducible
post-proxy association under the contract.

\subsection{Return objects and exact coefficient identity}

For event-session date $d_i$, let $P_i^L=C_{i,d_i-1}$ be the preceding
session close and let $P_i^E=C_{i,d_i+1}$ be the common following-session
close. Raw stock-price log returns in percentage points are
\begin{align}
 r_i^c &=100\log(P_i^E/P_i^c),\qquad c\in\{L,S,A\},\\
 d_i^{LS}&=100\log(P_i^S/P_i^L),\qquad
 d_i^{SA}=100\log(P_i^A/P_i^S).
 \label{eq:returns}
\end{align}
Log returns are used because adjacent intervals add exactly; ordinary simple
returns would instead compound. We write
$d_i^{LA}=d_i^{LS}+d_i^{SA}$. Event by event,
$r_i^L-r_i^S=d_i^{LS}$,
$r_i^S-r_i^A=d_i^{SA}$, and
$r_i^L-r_i^A=d_i^{LS}+d_i^{SA}$.

\paragraph{Paired OLS identity.}
Let $X$ be the same full-rank design matrix for all outcomes in
Eq.~\ref{eq:returns}. Let $\widehat\beta^{LS}$ and $\widehat\beta^{SA}$ denote
the corresponding same-design OLS coefficients for outcomes $d_i^{LS}$ and
$d_i^{SA}$. Then
\begin{equation}
 \widehat\beta^L-\widehat\beta^S=\widehat\beta^{LS},\quad
 \widehat\beta^S-\widehat\beta^A=\widehat\beta^{SA},\quad
 \widehat\beta^L-\widehat\beta^A
 =\widehat\beta^{LS}+\widehat\beta^{SA}.
 \label{eq:identity}
\end{equation}
OLS linearity gives Eq.~\ref{eq:identity} under the same estimator. This
diagnostic attributes a difference to the removed segment while holding
rows, representation, controls, fixed effects,
and endpoint fixed. Each contrast is thus localized to its removed return
interval.

Because moving only label start shortens the holding period, a separate
$r_i^{390}$ begins at $P_i^A$ and ends after exactly 390 XNYS trading minutes.
It tests duration and is not part of Eq.~\ref{eq:identity}.

\section{Earnings-Call Case Study}

  \begin{figure*}[!t]
  \centering
  \includegraphics[width=\textwidth]{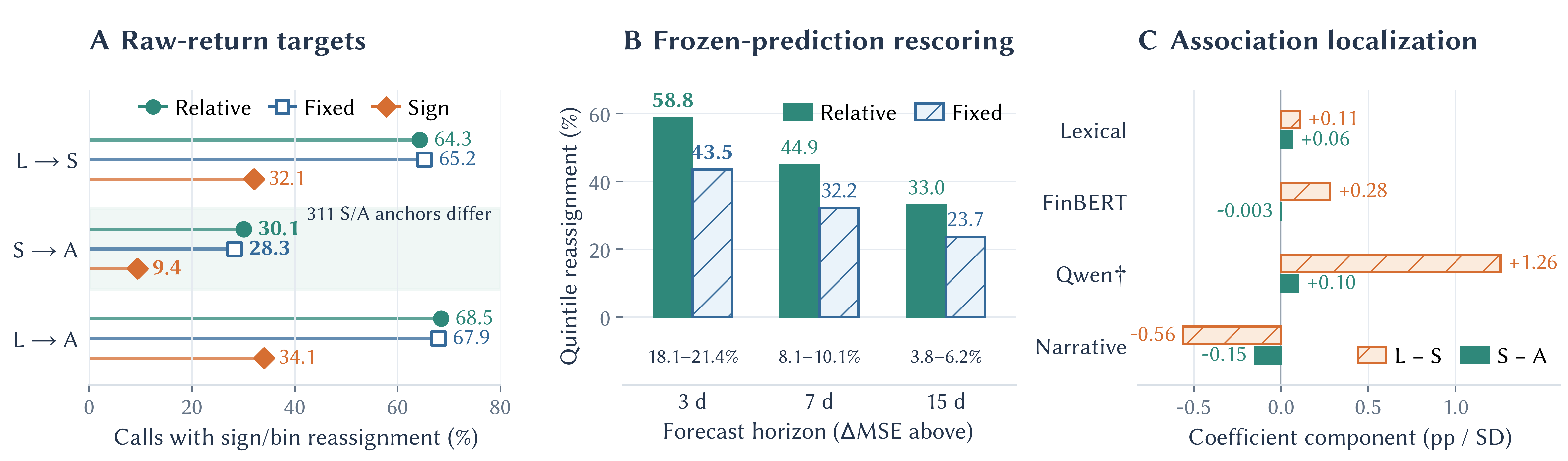}
  \caption{Target-clock sensitivity. (A) On 552 paired calls,
  $S\!\rightarrow\!A$ changes \SourceCompletionSignFlip{} of raw-return signs,
  \SourceCompletionFixedMove{} of fixed-$A$ quintiles, and
  \SourceCompletionRelativeMove{} of re-estimated relative quintiles. (B) A
  separate one-session perturbation of DEC
  before-market targets changes targets and four-stream MSE but not point
  rankings ($\tau=1$). (C) Exploratory $LS$ and $SA$ coefficient components;
  $\dagger$ marks the previously selected Qwen construct.}
  \Description{Panel A compares sign and quintile changes for the legacy to
  source-start, source-start to completion-proxy, and total clock adjustments.
  Panel B shows a separate one-session perturbation of DEC before-market targets:
  grouped relative and fixed-cutpoint quintile reassignment bars at three,
  seven, and fifteen days, with MSE changes below. Panel C shows the exploratory
  coefficient components removed over the first and second clock transitions
  for four representations.}
  \label{fig:results}
\end{figure*}

\subsection{Sample and clock construction}

The parent panel contains \PaperEventPanelN{} earnings calls for 30 large U.S.
firms from 2006--2024. The audit requires a usable source-page clock, archived
audio and sentence timing, fixed representation scores, and a complete
five-minute price path. The common sample contains \PaperNCalls{} calls from
\PaperNTickers{}
firms over 2018--2024, retaining \PaperRetention{} of the parent panel
(Table~\ref{tab:funnel}).

\begin{table}[t]
  \caption{Complete-case sample construction. Percentages are relative to the
  previous and parent panels.}
  \label{tab:funnel}
  \centering
  \small

\begin{tabular*}{\columnwidth}
  {@{\extracolsep{\fill}}lrrr@{}}
\toprule
Stage & Calls
      & \shortstack{Retained\\from prior}
      & \shortstack{Retained\\from parent} \\
\midrule
Parent event panel     & 2,094 & 100.0\% & 100.0\% \\
Usable source clock    & 755   & 36.1\%  & 36.1\%  \\
Audio-duration proxy   & 669   & 88.6\%  & 31.9\%  \\
\textbf{Complete common sample} & \textbf{552} & 82.5\% & 26.4\% \\
\bottomrule
\end{tabular*}

\end{table}

The deterministic parser processes 842 archived source pages. Six manually
verified start-time corrections and three verified no-change checks precede
sample freezing; most times remain documentary headers rather than verified
starts. Audio-duration construction yields 670 records, 669 linked to a usable
source clock.

Attrition is observably selective. The source-clock join removes 1,339 calls
and shifts the panel toward later years and larger firms; mean log market
capitalization moves by about one pooled SD. The price join removes 117 calls
from five firms (CRM, CVX, PG, TRV, and V). Only \PaperFirstYearN{} calls
remain in 2018, versus
\PaperLaterYearMinN{}--\PaperLaterYearMaxN{} per year thereafter. Inference
uses this common complete-case panel: clock evidence changes both the return
window and benchmark population. Price-window checks identified six split-like
discontinuities; none intersects an event window.

\paragraph{Worked example.}
We mechanically select the after-close call nearest the stratum's median audio
duration. For \WorkedTicker{} on \WorkedCallDate{}, the documentary header is
\WorkedCallStart{} ET, the audio proxy \WorkedAudioProxy{}, and stipulated
readiness proxy \WorkedReadyProxy{}. Both $S$ and $A$ map to the
\WorkedAnchor{} next-session close. The event-spanning, removed-$LS$, and
post-proxy returns are \WorkedLegacyReturn{}, \WorkedPreAnchorReturn{}, and
\WorkedPostProxyReturn{} pp, satisfying $r^L=d^{LS}+r^A$ to rounding. This
illustrates why the two proxy clocks yield the same return for an after-close call.

\subsection{Four retrospective representations}

Each Q\&A representation is aggregated to one call-level score and standardized
once on the common sample:
\begin{itemize}
  \item \textbf{Lexical tone}: answer-word-weighted counts from a
        project-curated, outcome-blind Loughran--McDonald-style dictionary
        subset.
  \item \textbf{FinBERT}: token-weighted
        $P(\mathrm{positive})-P(\mathrm{negative})$ from
        \texttt{ProsusAI/finbert}.
  \item \textbf{Qwen-derived answer valence}: a fixed Qwen3-4B prompt assigns
        each isolated Q\&A pair
        one label---positive, negative, neutral, or mixed. The call score is
        $(n_+-n_-)/n_{\mathrm{QA}}$; neutral and mixed answers contribute zero
        \cite{yang2025qwen3}.
  \item \textbf{Narrative revision}: for each question, E5 retrieves the
        nearest eligible prior question from the same firm. The exploratory
        score is mean cosine distance---one minus cosine similarity---between
        the current and prior answers, interpreted as a geometric proxy for
        narrative change \cite{li2026proxy}. Prior transcripts dated after the
        current call are ineligible. We use \texttt{intfloat/e5-large-v2}
        \cite{wang2022e5}.
\end{itemize}
The representations span lexical tone, finance sentiment, answer
classification, and within-firm narrative change. Qwen sees only an isolated
question and answer---not releases, guidance, consensus, prices, or returns.
Among 8,610 Q\&As, 48.62\% are positive, 48.19\% neutral, 2.54\% negative, and
0.65\% mixed. The score therefore mainly contrasts positive and neutral
answer valence.

Model revisions are fixed across the audit. Qwen uses greedy decoding, a limit
of 80 new tokens, and 3,072-token truncation. Its June 2026 run used a
2025-vintage checkpoint without an established historical knowledge cutoff;
the resulting scores are retrospective.

\subsection{Regression specification and inference}

For representation $z_i$ and outcome $r_i^k$, we estimate separate OLS models
\begin{equation}
 r_i^k=\alpha+\beta^k z_i+\gamma^{\mathsf T}W_i+\mu_{\mathrm{firm}}
 +\tau_{\mathrm{quarter}}+\lambda_{\mathrm{timing}}+\epsilon_i,
 \label{eq:regression}
\end{equation}

where $W_i$ contains standardized log Q\&A count and log answer-character
count; timing classes are before-open, intraday, and after-close. Every model
uses the same 552 rows and sparse controls to measure benchmark sensitivity
under a common design.
Here $\beta^k$ is measured in percentage points per signal SD; $\mu$, $\tau$,
and $\lambda$ are firm, calendar-quarter, and timing-class fixed effects.
Scores and controls remain unchanged across the three outcomes.

For target metrics, we report separate 5,000-draw 95\% percentile resampling
intervals from firm pairs-cluster and calendar-week
pairs-block resampling. Relative ranks and fixed-$A$ cutpoints are recomputed
within each draw. For exploratory association coefficients, the same two
one-way resampling schemes yield separate two-sided, null-centered approximate
$p$-values; the primary max-$p$ rule reports the larger of the firm- and
calendar-week values.
Holm adjustment \cite{holm1979multiple} covers four max-$p$ values;
Benjamini--Hochberg \cite{benjamini1995fdr} applies within the secondary
exact-390-minute family.

As a post-freeze sensitivity, we report the Cameron--Gelbach--Miller estimator
$V_{\mathrm{firm}}+V_{\mathrm{week}}-V_{\mathrm{firm}\times\mathrm{week}}$
\cite{cameron2011multiway}. It uses 24 firm clusters, 175 calendar-week
clusters, and 552 firm--week intersections (reference df 23). This asymptotic
check differs from the primary max-$p$ rule over separate one-way bootstraps.
Qwen and other constructs were inspected before family freezing; all
representation-level inference is therefore retrospective and exploratory.
The Holm family covers the four-representation $L\!\rightarrow\!A$
comparison. The subsequently added three-clock decomposition and DEC
fixed-cutpoint analysis are descriptive.

\paragraph{Implementation assistance.}
Claude and Codex assisted literature discovery, development and review of
audit scripts, and checks of result files and table consistency. All reported
estimates are computed by the audit scripts from the study data.

\section{Results}

\begin{table*}[t]
  \caption{Distinct target-clock sensitivities. $S\!\rightarrow\!A$ reassigns
  raw-return targets; separately, perturbing DEC before-market targets changes
  losses under frozen-prediction rescoring but not point rankings.}
  \label{tab:main}
  \centering
  \small
  \setlength{\tabcolsep}{3.0pt}
  \renewcommand{\arraystretch}{0.88}
\begin{minipage}[t]{0.485\textwidth}
\raggedright\textbf{Panel A. Three-clock raw-return sensitivity}
\par\vspace{0.5pt}
\begin{tabular*}{\linewidth}{@{\extracolsep{\fill}}lrrrr@{}}
\toprule
Metric & $L\!\to\!S$ & \textbf{$S\!\to\!A$} & $L\!\to\!A$ & After-close \\
\midrule
Calls & 552 & 552 & 552 & 206 \\
Spearman $\rho_s$ & 0.506 & 0.900 & 0.465 & 1.000 \\
Mean $|\Delta r|$ (pp) & 3.427 & 0.623 & 3.520 & 0.000 \\
Median $|\Delta r|$ (pp) & 2.661 & 0.179 & 2.587 & 0.000 \\
Sign flips (\%) & 32.1 & \textbf{9.4} & 34.1 & 0.0 \\
Relative moves (\%) & 64.3 & \textbf{30.1} & 68.5 & 6.3 \\
Fixed-$A$ moves (\%) & 65.2 & \textbf{28.3} & 67.9 & 0.0 \\
\bottomrule
\end{tabular*}
\end{minipage}\hfill
\begin{minipage}[t]{0.485\textwidth}
\raggedright\textbf{Panel B. External frozen-prediction sensitivity}
\par\vspace{0.5pt}
\begin{tabular*}{\linewidth}{@{\extracolsep{\fill}}lrrr@{}}
\toprule
Metric & \textbf{3 days} & 7 days & 15 days \\
\midrule
Events & 1,710 & 1,710 & 1,710 \\
Target $\rho_s$ & 0.645 & 0.822 & 0.902 \\
Relative moves (\%) & 58.8 & 44.9 & 33.0 \\
Fixed moves (\%) & 43.5 & 32.2 & 23.7 \\
Median rank shift & 12.5 & 8.8 & 6.1 \\
$\Delta$MSE range (\%) & \textbf{18.1--21.4} & 8.1--10.1 & 3.8--6.2 \\
$P$(same best), issuer/quarter & 0.67/0.72 & 0.95/0.83 & 0.86/0.80 \\
\bottomrule
\end{tabular*}
\end{minipage}

  \smallskip


  \parbox{\textwidth}{\footnotesize\textit{Notes.} Bold marks focal comparisons
and headline results, not statistical significance. Panel A recomputes relative
quintiles under each clock; fixed bins use $A$-clock cutpoints. Sign flips
exclude zeros. Equal after-close $S/A$ values can change relative bins through
the full-sample reference distribution. In Panel B, ranges span four prediction
streams; median rank
shifts are percentile points, and best-MSE retention uses issuer/quarter
resampling.}
\end{table*}

\begin{table*}[!t]
  \caption{Descriptive design sensitivities.}
  \label{tab:design-sensitivity}
  \centering
  \footnotesize
  \setlength{\tabcolsep}{3.0pt}
  \renewcommand{\arraystretch}{1.05}
\begin{tabular*}{\textwidth}{@{\extracolsep{\fill}}p{0.76in}p{1.02in}p{2.80in}p{2.10in}@{}}
\toprule
Dimension & Variation & Focal estimate or range & Design details \\
\midrule
\textbf{Allowance} $\ell$ & 0--120 min &
  Sign: 6.5\%--15.0\%; relative: 23.9\%--39.9\%; fixed-$A_{15}$: 21.0\%--35.9\% &
  Stipulated allowance scenarios. \\
Audio-duration $G$ & MP3 / alignment & 8 / 3 anchors move &
  Maximum coefficient change $<0.004$ pp/SD. \\
Endpoint $E$ & Following close / 390 min & $-0.233$ / $-0.226$ &
  Secondary BH $q=0.754$--$0.964$. \\
Dependence & Two-way CGM & $\Delta\beta=+1.355\;[+0.505,+2.204]$ &
  24 firms, 175 weeks; asymptotic sensitivity. \\
Concentration & Delete one group & Firm: 1.149--1.456; year: 1.119--1.540 &
  Delete-one-firm and delete-one-year refits. \\
\bottomrule
\end{tabular*}

  \smallskip

  \parbox{\textwidth}{\footnotesize\textit{Notes.} Fixed-$A_{15}$ uses the
  main 15-minute cutpoints throughout. The allowance grid varies the stipulated
  parameter; concentration checks refit the same specification after deleting
  one firm or year.}
\end{table*}

\subsection{Completion Alignment Reassigns Targets}

Panel A of Table~\ref{tab:main} and Figure~\ref{fig:results} separate two
operations that the original $L\!\rightarrow\!A$ comparison combined. Moving
from the previous close to the documentary source-start anchor
($L\!\rightarrow\!S$) gives Spearman
correlation \LegacySourceSpearman{}, a mean absolute return change of
\LegacySourceMeanAbs{} pp, \LegacySourceSignFlip{} sign changes, and
\LegacySourceRelativeMove{} relative-quintile reassignment. These quantities
describe conventional event-window realignment.

At the stipulated 15-minute baseline, moving from $S$ to $A$ changes targets
even after source-proxy alignment. The two
returns have Spearman correlation \SourceCompletionSpearman{}, but their
absolute difference averages \SourceCompletionMeanAbs{} pp and has median
\SourceCompletionMedianAbs{} pp. Raw-return sign changes for
\SourceCompletionSignFlip{} of calls. Under common $A$-clock cutpoints,
\SourceCompletionFixedMove{} change quintile; recomputing the reference
distribution under each clock yields \SourceCompletionRelativeMove{}. The
separate 5,000-draw 95\%
percentile resampling intervals under firm pairs-cluster and
calendar-week pairs-block resampling are, respectively, sign
$[\SourceCompletionSignFirmLow,\SourceCompletionSignFirmHigh]$/
$[\SourceCompletionSignWeekLow,\SourceCompletionSignWeekHigh]$, relative
$[\SourceCompletionRelativeFirmLow,\SourceCompletionRelativeFirmHigh]$/
$[\SourceCompletionRelativeWeekLow,\SourceCompletionRelativeWeekHigh]$, and
fixed $[\SourceCompletionFixedFirmLow,\SourceCompletionFixedFirmHigh]$/
$[\SourceCompletionFixedWeekLow,\SourceCompletionFixedWeekHigh]$. These
intervals quantify sampling variation in the reassignment rates.

The fixed-cutpoint comparison asks whether changing a call's return crosses
the same category thresholds. The relative comparison asks how its position
changes within the distribution defined by each clock. Reporting both
separates movement against a shared reference from redistribution within the
panel, making the categorical consequences of the clock adjustment explicit.

The 552-call rates are primary. $S$ and $A$ map to different anchors for
\SourceCompletionChangedAnchorN{}/552 calls
(\SourceCompletionChangedAnchorRate{}); within this price-anchor-different
subset, descriptive sign/relative/fixed rates are
\SourceCompletionAtRiskSignFlip{}/\SourceCompletionAtRiskRelativeMove{}/
\SourceCompletionAtRiskFixedMove{}. Among the
\SourceCompletionEqualAnchorN{} equal-anchor calls, numeric returns and
fixed-$A$ bins never change, whereas
\SourceCompletionEqualAnchorRelativeMove{} of calls change relative quintile only
because each clock uses a different full-sample reference distribution.

For all \AfterCloseSourceCompletionN{} after-close calls, both proxies map to
the same next-session bar, so $P^S=P^A$ and returns coincide. Fixed-$A$ bins
never change; the \AfterCloseSourceCompletionRelativeMove{} relative-bin
movement in Table~\ref{tab:main} comes only from different full-sample
reference distributions. The coincident anchors provide a mechanical check
of the timestamp-to-price mapping. These \AfterCloseSourceCompletionN{} calls are a subset of the
\SourceCompletionEqualAnchorN{} equal-anchor calls.

\subsection{Frozen Rescoring Changes Losses, Not Rankings}

On DEC, we rescore four frozen prediction streams on \DECCommonN{} common
events: TSMixer and three TMLP configurations (DEC, Random All, and Random
Ticker) from Yu et al.~\cite{yu2025identity}. The published target is the
log population standard deviation of simple daily returns over each horizon.
To isolate target-clock sensitivity from model adaptation, the day-indexed
counterfactual keeps predictions fixed, shifts only
\DECBeforeMarketN{} before-market targets by one session, and leaves
after-market targets unchanged. The after-market stratum is a mechanical
control ($\rho=1$, mean absolute change zero, and no fixed-cutpoint moves);
relative bins can still move through the cross-sectional distribution.

Panel B of Table~\ref{tab:main} shows that target sensitivity attenuates with
horizon. Because exactly the \DECBeforeMarketN{} before-market events shift,
numeric targets change for \DECNumericTargetChange{} of event--horizon rows. At
three days, relative/fixed quintile reassignment is
58.8\%/\DECThreeFixedQuintileMove{}, median absolute percentile displacement is
\DECThreeMedianRankDisplacement{} points, and MSE under this counterfactual is
18.1--21.4\% higher across the four frozen prediction streams. By 15 days these
values fall to
33.0\%/\DECFifteenFixedQuintileMove{},
\DECFifteenMedianRankDisplacement{} points, and 3.8--6.2\%.
Point rankings are unchanged under mean squared error (MSE), mean
absolute error (MAE), and Spearman correlation at every horizon (Kendall's
$\tau=1$). Issuer and calendar-quarter resampling further assess the
stability of these orderings.

Under issuer resampling, the probability of retaining the same best-MSE model
is \DECThreeSameBestIssuer{}, \DECSevenSameBestIssuer{}, and
\DECFifteenSameBestIssuer{} at 3, 7, and 15 days; calendar-quarter resampling
gives \DECThreeSameBestQuarter{}, \DECSevenSameBestQuarter{}, and
\DECFifteenSameBestQuarter{}. At three days, at least one model pair reverses in
\DECThreeAnyInversionIssuer{} of issuer resamples and
\DECThreeAnyInversionQuarter{} of calendar-quarter resamples. Target
perturbation, unchanged point rankings, and near-tie uncertainty coexist.

\subsection{Associations Localize before Source Start}

Table~\ref{tab:association-localization} localizes the retrospective
representation coefficients across the same three clocks. Fixed-prompt Qwen valence has
$\widehat\beta^L=\ValenceLegacyBeta{}$,
$\widehat\beta^S=\ValenceSourceBeta{}$, and
$\widehat\beta^A=\ValenceAvailableBeta{}$ pp per signal SD. Hence the total
$L\!\rightarrow\!A$ shift of \ValencePairedBeta{} decomposes into
\ValenceLegacySourceDelta{} over $L\!\rightarrow\!S$ and only
\ValenceSourceAvailableDelta{} over $S\!\rightarrow\!A$. Most of the
association shift is therefore localized before the documentary source-start
anchor, with a smaller component over the completion-proxy interval.

The original $L\!\rightarrow\!A$ 95\% firm- and calendar-week intervals are
$[\ValenceFirmCILow,\ValenceFirmCIHigh]$ and
$[\ValenceWeekCILow,\ValenceWeekCIHigh]$, with retrospective four-test Holm
$p=\ValenceHolmP$. Together with
Table~\ref{tab:association-localization}, these estimates show how the same
representation's association changes across clocks.

\subsection{Holding-Period and Design Sensitivities}

The common-endpoint audit moves start and preserves the additive identities;
the fixed-horizon audit holds the $A$ start and changes elapsed time.
Table~\ref{tab:design-sensitivity} varies allowance, audio-duration component,
dependence estimator, and group concentration.

Conditional on the same $A$ start, the following-close and exact-390-minute
Qwen point estimates are close
($\ValenceAvailableBeta{}$ and $\ValenceExactBeta{}$),
while the exact-horizon coefficients remain nonsignificant after multiplicity
correction. Across the 0--120-minute allowance grid,
sign/relative/fixed-$A_{15}$ move rates span
\AllowanceSignFlipMin{}--\AllowanceSignFlipMax{}/
\AllowanceRelativeMoveMin{}--\AllowanceRelativeMoveMax{}/
\AllowanceFixedAFifteenMoveMin{}--\AllowanceFixedAFifteenMoveMax{}.
The 15-minute baseline sign-flip rate of \SourceCompletionSignFlip{} lies
within this allowance-sensitivity envelope.

\section{Discussion and Availability Card}

\subsection{A practical benchmark specification}

Our Availability Card specializes BenchmarkCards
\cite{sokol2025benchmarkcards} to representation completion, clock evidence,
price anchors, endpoints, and permitted claims. Table~\ref{tab:card}
documents this audit; paired and fixed-duration tests separate label-start
from holding-period sensitivity.

The fields connect the measurement steps. The input boundary states what a
representation includes; the completion rule and allowance specify its timing
proxy; and the model record identifies the checkpoint. The anchor and endpoint
then define the return, while the sample and inference records describe the
observations used to estimate sensitivity. This organization allows another
study to retain the same reporting structure while choosing clocks appropriate
to its inputs and financial outcome.

\begin{table}[!t]
  \caption{Exploratory association localization.}
  \label{tab:association-localization}
  \centering
  \normalsize
  \setlength{\tabcolsep}{1.6pt}
  \renewcommand{\arraystretch}{0.92}
\begin{tabular*}{\columnwidth}{@{\extracolsep{\fill}}lrrrr@{}}
\toprule
Estimate & Lex. tone & FinBERT & Qwen val.$^{\dagger}$ & Narr. rev. \\
\midrule
$\beta_L$ & +0.100 & +0.196 & +1.122 & -0.579 \\
$\beta_S$ & -0.009 & -0.084 & -0.135 & -0.017 \\
$\beta_A$ & -0.073 & -0.082 & -0.233 & +0.134 \\
\addlinespace[1pt]
$\beta_L-\beta_S$ & +0.109 & +0.281 & +1.256 & -0.561 \\
$\beta_S-\beta_A$ & +0.064 & -0.003 & +0.098 & -0.151 \\
$\beta_L-\beta_A$ & +0.173 & +0.278 & +1.355 & -0.712 \\
\bottomrule
\end{tabular*}

  \vspace{2pt}
  \parbox{\columnwidth}{\footnotesize\raggedright
  \textit{Notes.} Coefficients are in percentage points (pp) per signal SD.
  Entries are retrospective point estimates.
  $\dagger$ marks the previously selected construct. Independently rounded
  components may differ from displayed totals by 0.001.}
\end{table}

\subsection{Interpreting the comparison}

Target reassignment and association localization address distinct questions.
The $S\!\rightarrow\!A$ comparison isolates completion-proxy alignment with
calls, scores, controls, and endpoints fixed, while most of the exploratory
Qwen shift occurs over $L\!\rightarrow\!S$. DEC separately shows that stable
rankings can coexist with changed losses under release-session realignment.

The baselines are measurement clocks applied to the same financial event.
$L$ captures the previous-close window; $S$ applies source-start dating; and
$A$ incorporates the audio-duration proxy and allowance. Reporting only
$L\!\rightarrow\!A$ would combine the first two adjustments. The paired
$S\!\rightarrow\!A$ comparison instead measures the incremental effect of
completion-proxy alignment. The four heterogeneous representations serve as
fixed measurement probes.

The appropriate window follows the research question. An event-response study
can use $L$ to capture price change around the disclosure, whereas an
evaluation based on the completed representation specifies a later input
boundary. Delaying label start also shortens the interval to a common
endpoint. The separate 390-trading-minute check retains the $A$ start and
fixes elapsed trading time. Together, these designs distinguish the choice
of forecast origin from the choice of evaluation horizon.

The three empirical layers examine target reassignment, association
localization, and frozen-prediction loss sensitivity. A changed quintile
records a change in the target assigned to a call; a coefficient contrast
locates the part of a fixed score's association removed by changing the clock;
and frozen rescoring measures the effect on reported predictive loss. These
comparisons explain how identical calls and model outputs can receive
different benchmark evaluations when the outcome clock changes.

Clock specification also clarifies what a model comparison measures. An early
partial transcript and a complete Q\&A representation contain different
information and may support different forecast origins. A common-anchor
comparison holds the forecast origin fixed; separate clocks instead preserve
each representation's declared availability boundary. Paired rescoring with
predictions fixed isolates the effect of target definition, whereas retraining
under each clock would additionally measure model adaptation. These designs
answer complementary questions and can be reported in sequence, distinguishing
differences in the evaluation task from differences in the predictor.

For example, an early-summary score and a complete-Q\&A score can be compared
from a common later anchor to study their relative predictive content.
Evaluating each from its own declared boundary instead compares two forecasting
tasks, each with a different information set and return interval. Keeping this
distinction explicit helps explain performance differences without treating
either window as the universal benchmark.

\paragraph{Implications for benchmark reporting.}
Predeclare the contract, freeze a joined sample, and report attrition.
Common-endpoint and fixed-duration tests distinguish label-start from
holding-period sensitivity. Representations with different readiness times
require a shared supported anchor or separately declared tasks. Recording the
input boundary, timestamp provenance, market calendar, checkpoint, allowance,
and endpoint makes these choices reproducible before model comparison.

\begin{table}[t]
  \caption{Availability Card for the earnings-call audit.}
  \label{tab:card}
  \centering
  \small
  \renewcommand{\arraystretch}{1.0}
  \begin{tabular}{@{}p{0.6in}p{2.5in}@{}}
    \toprule
    Field & This audit \\
    \midrule
    Claim & Proxy-aligned three-clock target and association sensitivity. \\
    Input boundary & Complete spoken Q\&A. \\
    Readiness $(G,\ell,M)$ & Reconstructed audio-duration proxy plus 15 minutes;
      0--120 minutes is the sensitivity grid. Transcript and model-inference
      completion are unobserved. \\
    Model clock & Retrospective checkpoints; Qwen inference ran in June 2026
      with a 2025-vintage checkpoint. \\
    Label start & $L$: previous close; $S$: close of the first bar starting
      after the documentary source-start proxy; $A$: the corresponding rule
      after the readiness proxy. These are label prices, not execution prices. \\
    Endpoint & Following-session close; secondary endpoint after exactly 390
      trading minutes. \\
    Price data & Archived raw intraday stock-price series; upstream vendor,
      timezone declaration, and adjustment/total-return semantics are not
      independently certified. \\
    Sample & 552 complete calls from 24 firms after clock, audio, score, and
      price joins. \\
    Statistical\newline inference & Common-row OLS; separate 5,000-draw firm and
      calendar-week resampling; post-freeze CGM sensitivity; Holm and BH
      within-family adjustments; representation analysis is exploratory. \\
    \bottomrule
  \end{tabular}
\end{table}

\section{Conclusion}

The Availability Contract links a financial NLP benchmark's input boundary,
timing evidence, and outcome clock. The paired three-clock audit separates
event dating from completion-proxy alignment while preserving a common sample
and endpoint. At the stipulated 15-minute baseline, $S\!\rightarrow\!A$
changes raw-return signs for \SourceCompletionSignFlip{} of calls and
fixed-$A$/relative assignments for
\SourceCompletionFixedMove{}/\SourceCompletionRelativeMove{}. Independently,
the DEC release-session analysis raises three-day MSE by 18.1--21.4\% under
frozen predictions while leaving point rankings unchanged. The contract and
Availability Card make outcome-clock choices explicit and provide a practical
basis for reporting target and loss sensitivity alongside model comparisons.

\clearpage
\bibliographystyle{ACM-Reference-Format}
\bibliography{references}

\end{document}